\documentclass[aps,preprint,superscriptaddress,groupedaddress]{revtex4}  
\usepackage{color,graphicx}  
\usepackage{dcolumn}   
\usepackage{bm}        
\usepackage{amssymb}   
\usepackage{amsmath}
\DeclareGraphicsExtensions{.pdf,.pgn,.jpg, .eps}
\usepackage{subfig}

\begin{document}

\widetext

\title{A Simplified Self-Consistent Probabilities Framework to Characterize Percolation Phenomena on Interdependent Networks : An Overview}

\date{\today}

\author{Ling Feng}
\affiliation{Complex Systems Group, Institute of High Performance Computing, Agency for Science Technology and Research, Singapore 138632}

\author{Christopher Pineda Monterola}
\affiliation{Complex Systems Group, Institute of High Performance Computing, Agency for Science Technology and Research, Singapore 138632}
\author{Yanqing Hu}
\email[To whom correspondence should be addressed; ]{yanqing.hu.sc@gmail.com}
\affiliation{School of Mathematics, Southwest  Jiaotong University, Chengdu 610031, China}

\begin{abstract}
Interdependent networks are ubiquitous in our society, ranging from infrastructure to economics, and the study of their cascading behaviors using percolation theory has attracted much attention in the recent years. To analyze the percolation phenomena of these systems, different mathematical frameworks have been proposed including generating functions, eigenvalues among some others. These different frameworks approach the phase transition behaviors from different angles, and have been very successful in shaping the different quantities of interest including critical threshold, size of the giant component, order of phase transition and the dynamics of cascading. These methods also vary in their mathematical complexity in dealing with interdependent networks that have additional complexity in terms of the correlation among different layers of networks or links. In this work, we review a particular approach of simple self-consistent probability equations, and illustrate that it can greatly simplify the mathematical analysis for systems ranging from single layer network to various different interdependent networks. We give an overview on the detailed framework to study the nature of the critical phase transition, value of the critical threshold and size of the giant component for these different systems.
\end{abstract}


\maketitle

\section{Introduction}

Systems consisting of multiple inter-connected networks with different types of links have received enormous attention in the recent years \cite{SergeyNature2010,KurantPRLlayer2006,MuchaScienceMultiplex2010,RenaudPNASMultiplex2010,MorrisPRL2012CoupleSpatial,RaissaPNAS2012,YanqingNP2014,BaxterPRL2012,YanqingPRX2014}, due to its ubiquitous applications in complex systems. Such networks appear in the literature known as interdependent networks or multiplex networks. Studies have shown that interdependent networks show distinct percolation/ phase transition behaviors from single networks. In particular, an interdependent network is more vulnerable to random attacks \cite{gao2012robustness}. As many real world infrastructure networks can be classified into interdependent networks \cite{peerenboom2001,StevenIEEE2001,rinaldi2004,panzieri2008}, the understanding of their robustness carries great practical significance.

In a network consisting of links and nodes, one of the most important quantities used to analyze its robustness is the size of the giant component, which is defined as the largest set of nodes that are connected with each other. When a network is under attack, i.e. a fraction $1-p$ of nodes (or links) are removed, the size of the largest cluster shrinks. Usually its size is a finite fraction of the total number of nodes in the network, unless more than a certain fraction $1-p_c$ of nodes are removed - then the largest cluster (as known as the giant component) disappears and all of the clusters become  negligibly small. This phase is associated with the disintegration of the network. Hence the size $\mu^{\infty}$ of the giant component serves as an order parameter that is very useful in studying the phase transition behaviors and the robustness of the network structure.

 One of the original works in \cite{SergeyNature2010} provided a precise and powerful analytical solution to the phase transition behaviors. In their mathematical analysis, recursive mapping was used to track the percolation process in each stage of cascading failures. In some systems where correlations exist in dependency links\cite{YanqingNP2014,YanqingPRX2014,BaxterPRL2012,YanqingPRE2013,LidiaPRETriplePoint2013}, this method could lead to very complicated formulations and not always easy to solve.
To study different network constructions, some of the other studies used different methods to achieve a relatively simpler  analytical framework. In particular, the works in \cite{SeungEPL2012,MendosPRLKcore2006,MendesBootstrap2010,BaxterPRL2012,YanqingNP2014,GohPRECorr2014} used self-consistent equations of the converging probabilities to have an alternative approach to analyze for the critical behaviors on certain types of interdependent networks. Some of these methods can be extended to other scenarios.

        In this paper, we illustrate the use of one particular technique based on self-consistent probabilities\cite{MendosPRLKcore2006,MendesBootstrap2010,SeungEPL2012,BaxterPRL2012}, and demonstrate that it could be applied to a wide variety of different interdependent networks with minimum simplicity through surveying the literature in this field. This method  focuses on the recursive representation of two central quantities defined as the probabilities of finding a link/node in the giant component. It is able to give a set of straight forward self-consistent equations describing the percolation behaviors without going through the cascading process \cite{SergeyNature2010}, and also deal with many correlated systems with simpler mathematics formulations.

First we will illustrate the framework through the example of the single layer network. Next we extend it to multi-layer networks without degree degree correlations. Following that we extend the analysis to more complicated scenarios of partially correlated networks and degree-degree correlated networks. More complications are added to the case when multiple dependency links per node is introduced together with correlations, as well as single network with different types of links, also known as multiplex networks.

\section{Single Layer Network}

\begin{figure}[h]
 \includegraphics[width=0.3\linewidth]{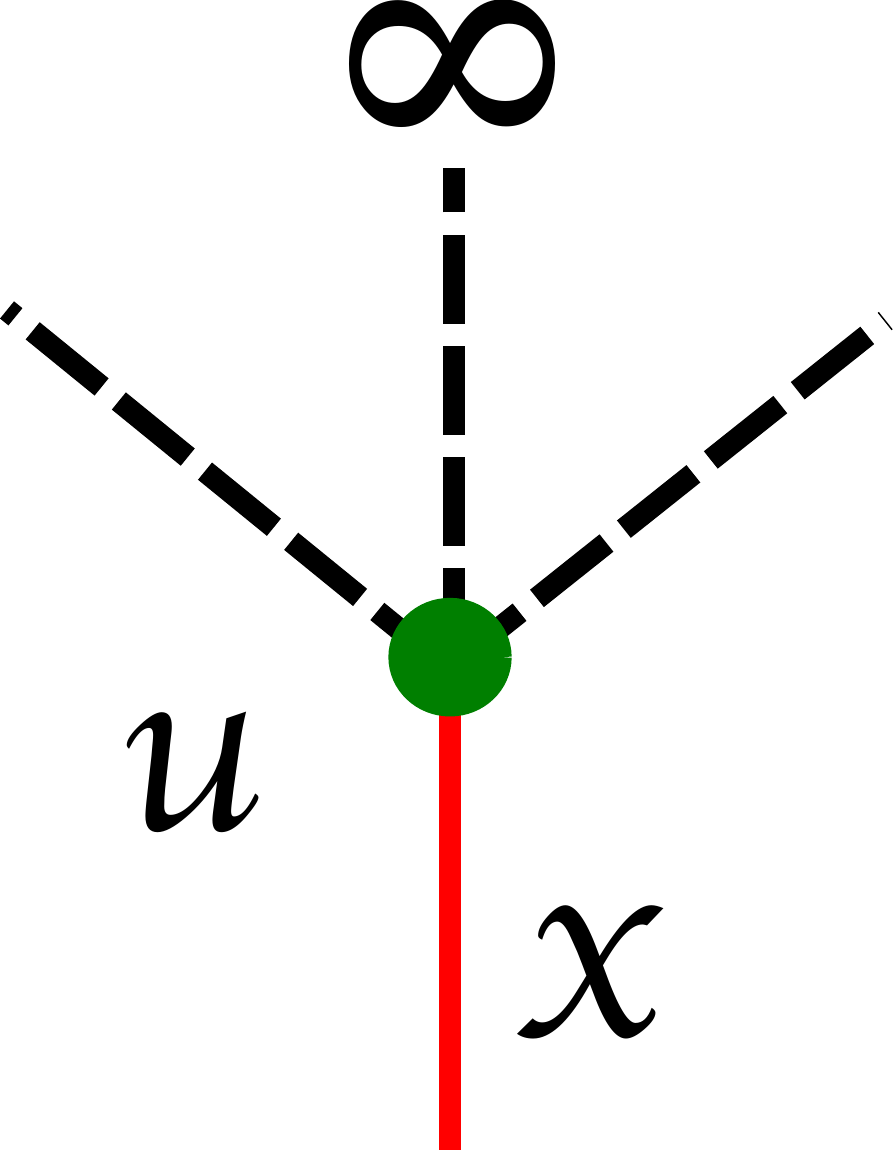}
 \caption{Definition of $x$. A link (red color) is chosen and a node $u$ (in green) is found. From the three outgoing links (dashed lines) of node A, one of them leads to  to the giant component (represented by the symbol $\infty$). Since at least one of the three outgoing links of $u$ leads to the giant component, the red link leads to the giant component. And we define the probability of finding such a red link as $x$.}
\label{FIG:DefineX}
\end{figure}

The classic site percolation problem in a random network \cite{NewmanPRL2000,Shlomo2000,NewmanPRE2001,CohenPREpcScalefree2002} gives rich phase transition phenomena for various networks structures. In the simplest case, we consider a random network without any correlations, and its degree distribution $P(k)$ fully captures its structural property. We start by introducing a key quantity $x$ in the system; This $x$ will be similarly defined throughout this work and plays a central role in the mathematical analysis. If we randomly choose a link from the network and travel along one direction of the link, there is a probability $x$ it would  reach the giant component of the network, and probability $(1-x)$ it will not.  (See Figure~\ref{FIG:DefineX} for illustration).

Suppose we randomly choose a link, and find an arbitrary node $u$ by following this link in an arbitrary direction. The probability that the node $u$ has degree $k'$ is 

\begin{equation}
\frac{P(k')k'}{\sum_k P(k)k}= \frac{P(k')k'}{\langle k \rangle}.
\end{equation}

For this node $u$ to be part of the giant component, at least one of its other $k-1$ out-going links (other than the link we first picked) leads to the giant component. By calculating this probability, we can write out the self-consistent equation for $x$:

\begin{equation}
x=\sum_k \frac{P(k)k}{\langle k \rangle} \cdot [1-(1-x)^{k-1}],
\label{En:single_x}
\end{equation}
where $1-(1-x)^{k-1}$ is the probability that at least one of the other $k-1$ links of node $u$ lead to the giant component, and $\frac{P(k)k}{\langle k \rangle}$ is the probability that the node $u$ has degree $k$.

Therefore, for a randomly chosen node $u$, the probability that it is in the giant component is equal to the probability that at least one of its $k$ links leads to the giant component. Thus we have:
\begin{equation}
\mu^{\infty}=\sum_k P(k)\cdot [1-(1-x)^k],
\label{En:single_mu}
\end{equation}
where $1-(1-x)^k$ is the probability that none of the $k$ links of node $u$ leads to the giant component, and $P(k)$ is the probability that node $u$ has degree $k$. It is worth noting that $\mu^{\infty}$ is also the normalized size of the giant component, i.e. the fraction of nodes in the giant component. The above equations exactly equal to the results obtained by M. E. J. Newman and et.al in \cite{NewmanPRE2001}.

In the network percolation problem, when a fraction of $1-p$ nodes are randomly removed from the network \cite{NewmanPRE2001,CohenPREpcScalefree2002}, i.e. there is a fraction of $p$ node remaining, we could apply the previous equations with slight modifications. 
Assuming that the links of the removed nodes are still present on the network, the probability that a randomly selected link leads to the giant component is the same as before, given by $\sum_k \frac{P(k)k}{\langle k \rangle} \cdot [1-(1-x)^{k-1}]$. But since only a fraction $p$ of the nodes remain in the network, by calculating the probability that the randomly chosen link does not lead to the giant component, the self-consistent equation of $x$ in En~(\ref{En:single_x}) becomes:

\begin{equation}
x=p\cdot\sum_k \frac{P(k)k}{\langle k \rangle} \cdot [1-(1-x)^{k-1}],
\label{En:singleP_x}
\end{equation}
where $1-(1-x)^{k-1}$ is the probability that at least one of the $k-1$ outgoing links of node $u$ leads to the giant component, and $\frac{P(k)k}{\langle k \rangle}$ is the probability that $u$ has degree $k$, same as before. The additional variable $p$ in front is due to the fact that only a fraction of $p$ nodes remain in the network after removing $1-p$ nodes.

Similarly, the probability that a randomly selected node is in the giant component is:

\begin{equation}
\mu^{\infty}=p\cdot \sum_k P(k)\cdot [1-(1-x)^k].
\label{En:singleP_mu}
\end{equation}

It is known that in a single network, we usually only have second order phase transitions, such that there is no giant component when  $p$ is smaller than a critical probability $p_c$. Above the threshold giant component appears and its size increases continuously from 0 with increasing $p$.
This means when $p\rightarrow p_c^{II}$, we have $x\rightarrow0$ and $\mu\rightarrow0$. When $x\rightarrow0$, by taking Taylor expansion of En~(\ref{En:singleP_x}), we obtain:

\begin{equation}
x = p_c^{II} \cdot  \sum_k \frac{P(k)k}{\langle k \rangle} \cdot (k-1)x+o(x),
\end{equation}
which leads to
\begin{equation}
p_c^{II} = \frac{\langle k \rangle}{\langle k(k-1)\rangle}
\label{En:single_pc}
\end{equation}

For Erdos-Renyi network, En~(\ref{En:single_pc}) yields $p_c^{II} = 1/ \langle k \rangle$, in agreement with the known result.  For scale free network with $\gamma<3$, $\langle k^2 \rangle$ diverges, thus we would obtain $p_c=0$, also in agreement with the known result.

The above system is based on node percolation, in which nodes are randomly removed until the giant component disintegrates. An alternative scenario is link (bond) percolation, in which links are randomly removed from the network. In this case, we still have the same definition for $x$, and its equation remains the same as En~\ref{En:singleP_x}, because a randomly selected link has probability $p$ to still remain in the network after removing a fraction of $1-p$ links. The only difference is in En. \ref{En:singleP_mu}, in which we need to remove $p$ on the right side:
\begin{equation}
\mu^{\infty}=\sum_k P(k)\cdot [1-(1-x)^k].
\label{En:singleP_mu'}
\end{equation}

This is due to the fact that all of the nodes remain in the network in link percolation, unlike the cases of node percolation that only a fraction of $p$ remains. Hence we would obtain the same $p_c^{II}$ value for both node and link percolation, but different $\mu^{\infty}$ values.

Before we proceed to interdependent networks, it is worth mentioning that other than second order phase transition mentioned above, there could also be first order phase transition, and the critical threshold can be labelled as $p_c^{I}$. For such phase transitions,when $p<p_c^{I}$ the size of the largest cluster is 0, and abruptly jumps to a non-zero value at $p=p_c^{I}$. We shall see more of such examples later.

\section{Multi-layer Interdependent Network}

\subsection{Two Layer Interdependent Network}

In the original work of Ref~\cite{SergeyNature2010}, generating functions was used to study the phase transitions in the two layer interdependent network. The system consists of two networks A and B, with degree distributions $P_A(k)$ and $P_B(k)$ respectively. Both networks A and B have N nodes, and each node in A is linked with exactly one node in B by a {\it dependency link}, and vice versa. The dependency link is different from the connectivity links within each network, in the way that once a node on one end of the dependency link is removed, the other node on the other network is also removed. This corresponds to the case where the failure of a power plant in the grid network will render the connected computer system to shut down due to the unavailability of electricity. Also, any node outside the giant cluster of its own network would fail since it is disconnected with the majority of the other nodes. In the defined mutually connected giant component (MCGC), every node is in the  giant component connected via the connectivity links in its own network, and its dependent node is in the giant component of the other network as well. Thus the MCGC is a steady state of the remaining network, such that no further cascading of failures would happen.


Here we present a simple method to study the phase transition behaviors using the formulation extended from the previous section. Following the definition in En~(\ref{En:single_x}), we define $x$ as the probability that a randomly chosen link in network A leads to the giant component. Analogously  the probability that a randomly chosen link in network B leads to the giant component is $y$.


\begin{figure}[h]
 \includegraphics[width=0.6\linewidth]{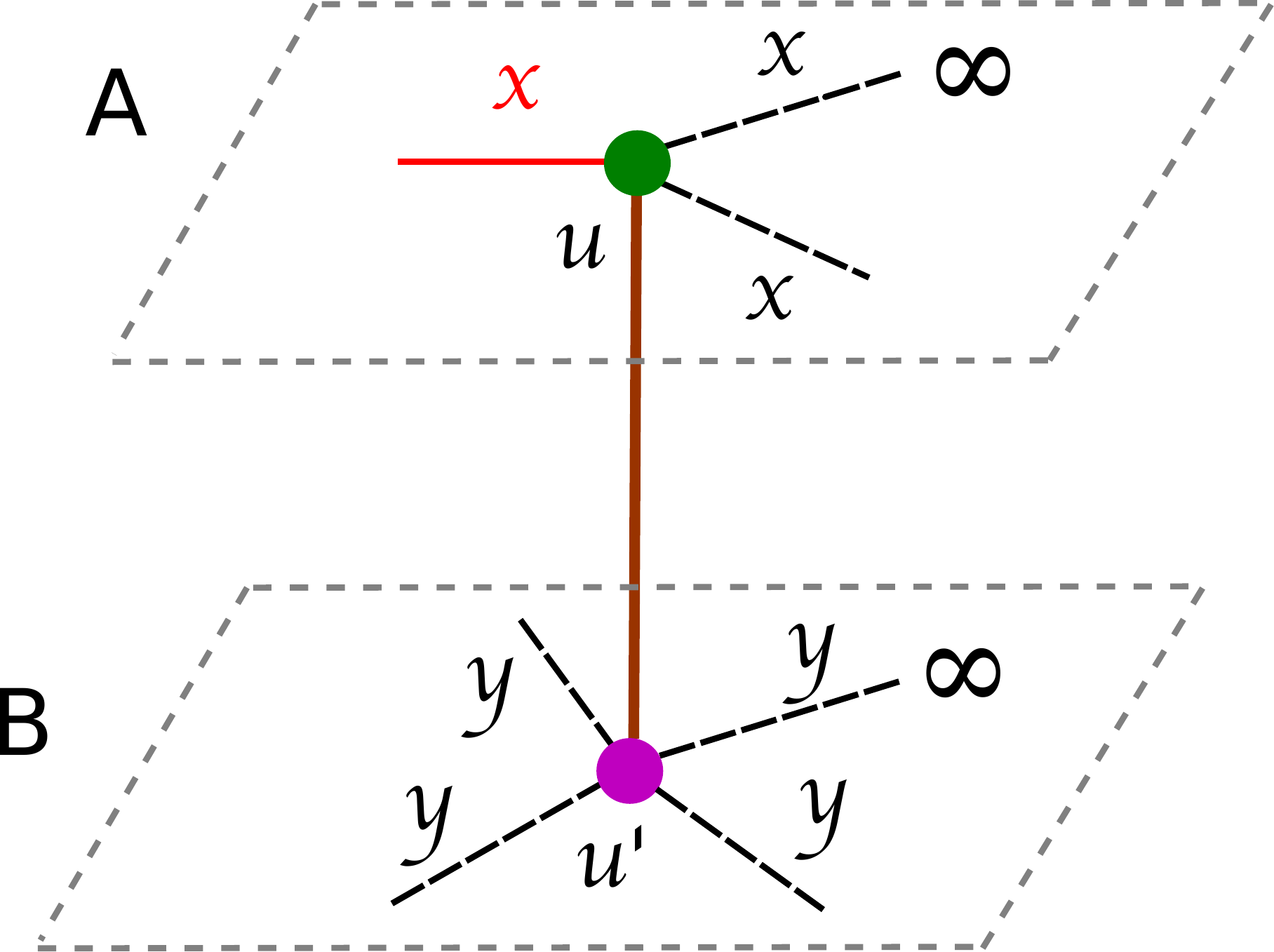}
 \caption{ Definition of $x$ and $y$ in interdependent network. Two networks (A and B) both with $N$ nodes are connected by dependency links, with one-to-one matching. Here node $u$ is connected with $u'$ via an interdependent link. 
  A link (red solid line) in the  network A is chosen and a node $u$ (in green) is found following the link. Out of the two outgoing connectivity links (black dashed lines) of $u$, one leads to the giant component. Node $u'$ in the lower network is connected with $u$ via the dependency link (solid brown line), and it is also connected with the giant component via a connectivity link in network B. Since both node $u$ and $u'$ are connected to the giant component, we can be sure that the initial red link leads to the mutually connected giant component (MCGC).  In network A, we define the probability of finding a connectivity link leading to the  MCGC as $x$.
  Similarly in network B, we define the probability of finding a connectivity link leading to the MCGC as $y$.}
  \label{FIG:DefineDoubleX}
  \end{figure}

If a randomly chosen link in A leads to a node with degree $k$, the node is in the MCGC only if at least one of its other $k-1$ links lead to the giant component, and its dependent node in network B is also in the MCGC. Otherwise, this link will be not be in the mutually connected giant component, and be eventually deleted according to Ref~\cite{SergeyNature2010} (see Figure~\ref{FIG:DefineDoubleX} for a detailed illustration).
Therefore by calculating the probability that a randomly chosen link in A leads to the MCGC, we would obtain:

 \begin{equation}
 x=\sum_k \frac{P_A(k)k}{\langle k_A \rangle}[1-(1-x)^{k-1}] \cdot \sum_{k'} P_B(k')[1- (1-y)^{k'}],
 \label{En:double_x}
 \end{equation}
 where $\sum_k \frac{P_A(k)k}{\langle k_A \rangle}[1-(1-x)^{k-1}]$ is the probability that at least one of node $u$'s other $k-1$ connectivity links in network A leads to MCGC, and $\sum_{k'} P_B(k')[1-(1-y)^{k'}]$ is the probability that at least one of the $k'$ connectivity links of the dependent node $u'$ in network B leads to the MCGC.

Similarly we would have the probability that a randomly chosen link in B leads to the MCGC:
  \begin{equation}
 y=\sum_k \frac{P_B(k)k}{\langle k_B \rangle}[1-(1-y)^{k-1}]\cdot \sum_{k'} P_A(k')[1-(1-x)^{k'}].
 \end{equation}
 
Consequently, the probability that a randomly chosen node (either in network A or B) is in the MCGC is:
 \begin{equation}
 \mu^{\infty}=\sum_k P_A(k)[1-(1-x)^k] \cdot \sum_{k'} P_B(k')[1-(1-y)^{k'}],
 \label{En:double_mu}
 \end{equation}
 which again is the normalized size of the mutually giant component. Note that we do not distinguish this value in different networks, because there is a one-to-one matching between nodes in A and B, so that $\mu^{\infty}$ is identical for both networks.

When we randomly remove $1-p$ fraction of nodes from network A, there is only $p$ fraction of nodes left in A. Hence out of the original probability $x$ that a randomly selected link leads to the MCGC, only a fraction of $p$ nodes are actually remaining. It is easy to write down the new expression for $x$ as:

 \begin{equation}
 x=p \cdot \sum_k \frac{P_A(k)k}{\langle k_A \rangle}[1-(1-x)^{k-1}] \cdot \sum_{k'} P_B(k')[1- (1-y)^{k'}].
  \label{STX}
 \end{equation}

Analogously, the equation for $y$ is

 \begin{equation}
 y=p \cdot \sum_k \frac{P_B(k)k}{\langle k_B \rangle}[1-(1-y)^{k-1}]\cdot \sum_{k'} P_A(k')[1-(1-x)^{k'}].
  \label{STY}
 \end{equation}
 
 At last, we arrive at the equation of $\mu^\infty$, which is the probability that a randomly selected node in A (or B) is in the MCGC:
 
 \begin{equation}
 \mu^{\infty}=p\cdot \sum_k P_A(k)[1-(1-x)^k] \cdot \sum_{k'} P_B(k')[1-(1-y)^{k'}].
 \label{STMU}
 \end{equation} 
which is also the normalized size of the MCGC.

\bigskip

In principle, Eqs. \ref{STX} and \ref {STY}  can be transformed into 
 \begin{equation}
 x=F_1(p, y),
 \end{equation}
and
 \begin{equation}
 y=F_2(p, x).
 \end{equation}
If we cannot get the explicit formula as above, the numerical computation always can be employed succesfully.
 
 Usually, the phase transition for the above system is of first order at the critical point $p_c^I$ (example given below). Therefore, at $p=p_c^I$, the two functions $x=F_1(p_c^I, y)$ and $y=F_2(p_c^I, x)$  meet tangentially with each other:
 
  \begin{equation}
 \frac{\partial F_1(p_c^I,y)}{\partial y} \cdot \frac{\partial F_2(p_c^I,x)}{\partial x}=1.
 \label{TA1}
 \end{equation}

 For the first order phase transition, at the the critical point $p_c^{I}$, the giant component is not 0, implying that we cannot employ Taylor expansion to simplify Eqs. \ref{STX} and \ref {STY}, but have to solve the polynomial equations directly. It could be very difficult to obtain the explicit formula for $p_c^{I}$ except the most simple distributions, but numerical methods are possible.  
 
 \bigskip

{\it Example with random regular network}

\bigskip

 For a simple example, we assume both network A and B are random regular networks with $P_A(3)=P_B(3)=1$. It means every node in both networks have degree 3, and the nodes are randomly connected.
 The above Eqs. \ref{STX}, \ref {STY} and \ref{STMU} then becomes:

 \begin{align}
    \label{ExampleRRX0}
 x=p [1-(1-x)^2][1- (1-y)^3],\\
   \label{ExampleRRY0}
    y=p [1-(1-y)^2][1-(1-x)^3],\\
  \mu^{\infty}=p [1-(1-x)^3][1-(1-y)^3].
 \end{align}

If $x\neq 0$ and $y\neq 0$, further simplification gives
 \begin{align}
     \label{ExampleRRX}
x=F_1(p, y)=2-\frac{1}{p[1-(1-y)^3]},
\\
   \label{ExampleRRY}
y=F_2(p, x)=2-\frac{1}{p[1-(1-x)^3]}.
 \end{align}

Hence the requirement for $p_c^I$ of equation \ref{TA1} can be written explicitly as
 \begin{equation}
\frac{3(1-y)^2}{p_c^I[1-(1-y)^3]^2}\cdot\frac{3(1-x)^2}{p_c^I[1-(1-x)^3]^2}=1.
\label{ExamplePC}
 \end{equation}
Solving the above three equations gives us $x=y\approx 0.5446$, $p_c^I\approx 0.7588$ and consequently the mutual giant component size  $\mu^{\infty}\approx 0.6329$.
 
\begin{figure}
\includegraphics[scale=0.4]{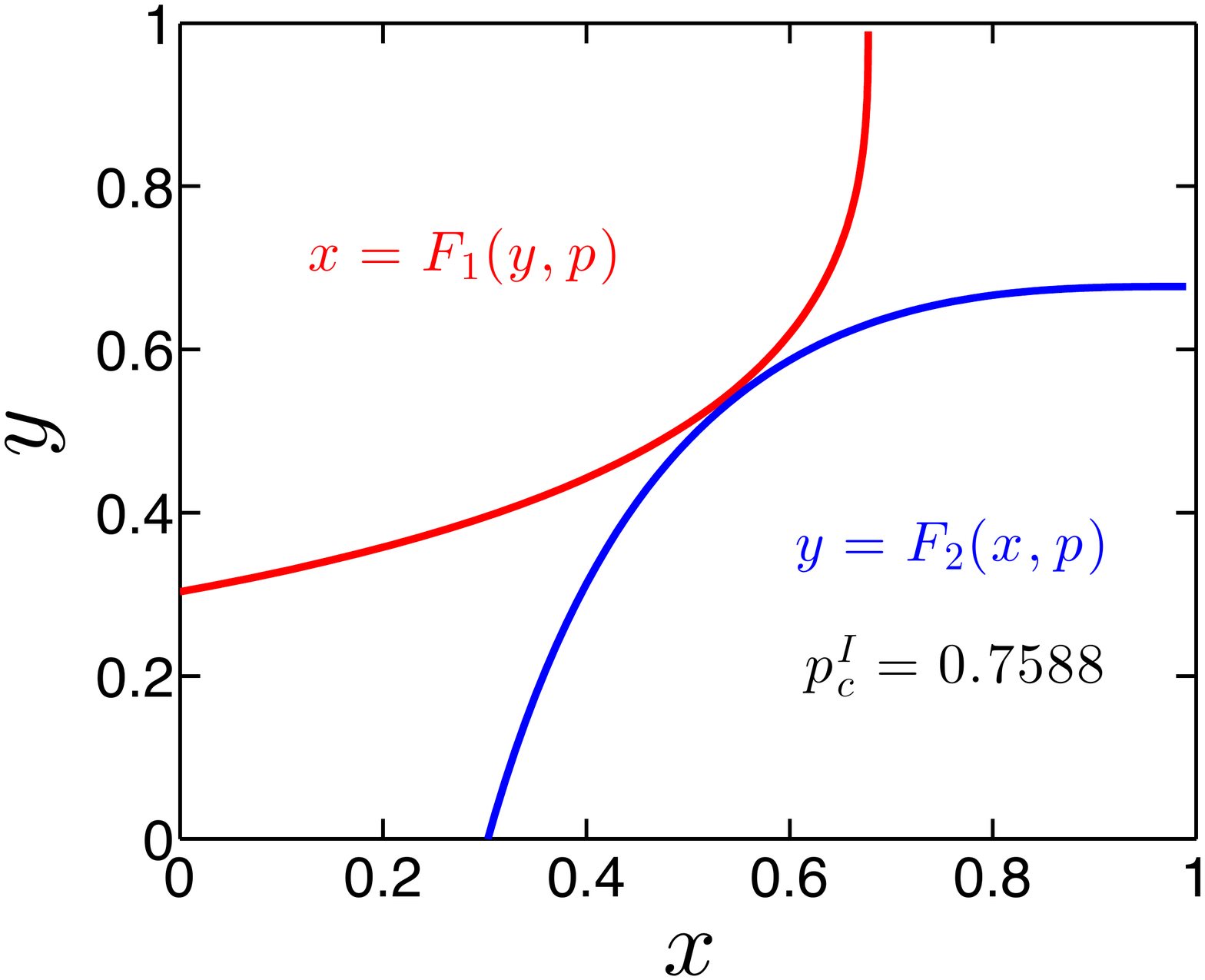}
\caption{\label{tangential} Solving for critical value of $p_c^I$ for random regular interdependent network with degree 3. The two curves represent  Eqs. \ref{ExampleRRX} and \ref{ExampleRRY}. At $p=p_c^I\approx 0.7588$, the curves from equations \ref{ExampleRRX} and \ref{ExampleRRY} touches each other at $x=y\approx 0.5446$, where the slope of the two curves are equal.}
\label{FIG:PC1}
\end{figure}

\begin{figure}
\includegraphics[scale=0.4]{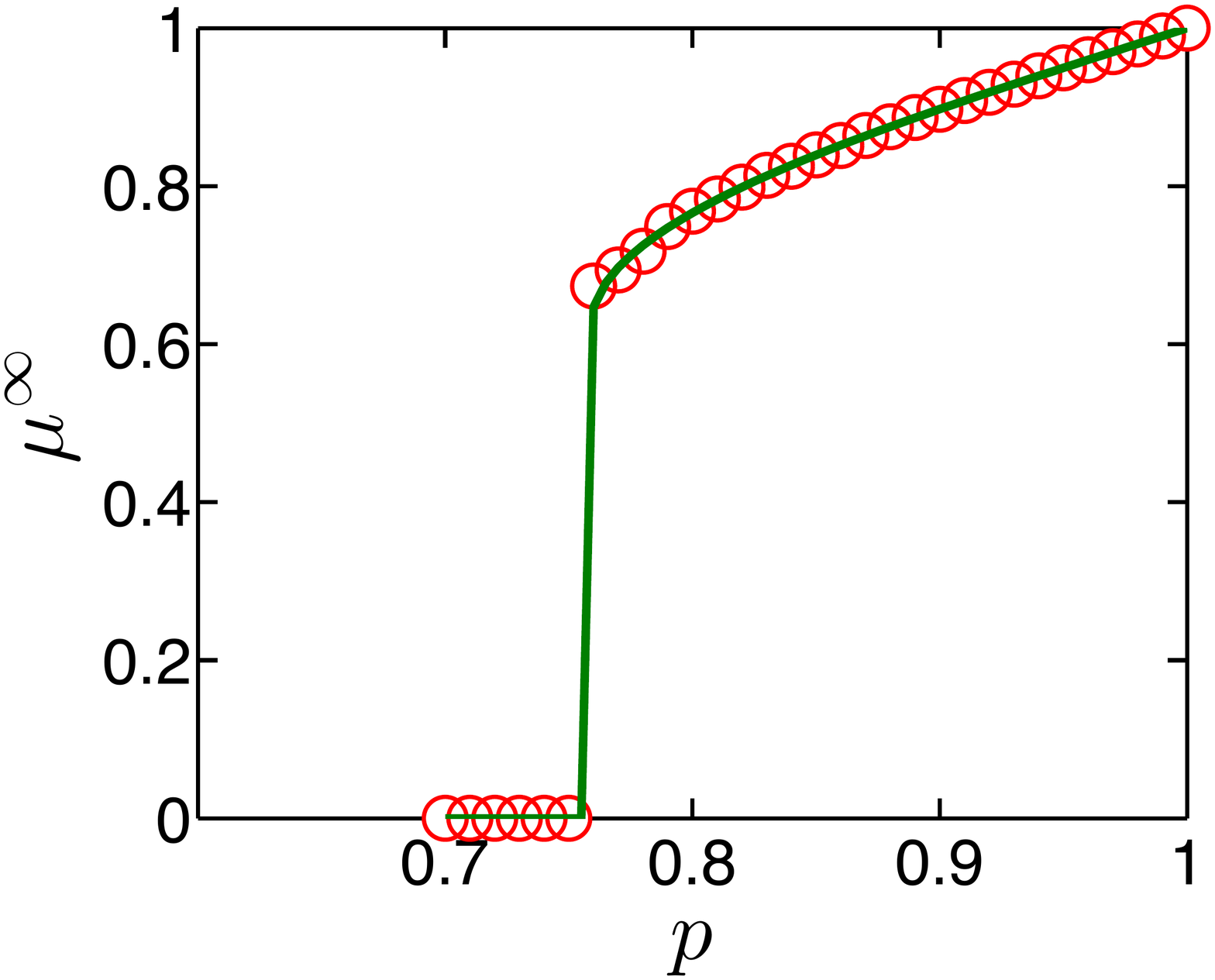}
\caption{First order phase transition on random regular interdependent network with degree 3.  The red circles represent simulation result and the green line is the theoretical result from En~\ref{ExamplePC}. $\mu^{\infty}$ jumps from 0 to 0.6329 at $p=p_c^I=0.7588$, representing a first order phase transition.}
\label{RRPCI}
\end{figure}

The tangential requirement in equation \ref{TA1} is presented in Figure~\ref{FIG:PC1}. When $p=p_c^I\approx 0.7588$, the curves from equations \ref{ExampleRRX} and \ref{ExampleRRY} touch each other at $x=y\approx 0.5446$, where the slope of the two curves are equal; When $p<p_c^I$, the two curves do not touch each other, and we only have the trivial solution of $x=y=0$ from equations \ref{ExampleRRX0} and \ref{ExampleRRY0}. This abrupt change in the size of the giant cluster corresponds to the first order transition, in which $\mu^{\infty}$ changes from 0 to 0.6329 abruptly at $p=p_c^I$. This is illustrated through simulation results in Fig~\ref{RRPCI}.

\subsection{N-layer interdependent network}

The case of N-layer interdependent networks \cite{JianxiPRE2013,JianxiPRL2011,JianxiNatPhy2011,gao2012robustness,BaxterPRL2012,BianconiPRE2014,YanqingNP2014} is an extension of the two layer scenario. Assuming there are N networks of equal number of nodes, and each node in a network is randomly connected with one and only one node in every other networks. Extending from En~(\ref{En:double_x}), we obtain the probability that a randomly chosen link in network $i$ leading the the MCGC as:

 \begin{equation}
 x_i=\sum_{k_i} \frac{P_i(k_i)k_i}{\langle k_i \rangle}[1-(1-x_i)^{k_i-1}] \prod_{j\neq i} \{\sum_{k_j} P_j(k_j)[1- (1-x_j)^{k_j}]\},
\label{En:N_x}
 \end{equation}
 where $\frac{P_i(k_i)k_i}{\langle k_i \rangle}$ is the probability that a randomly chosen link  in network $i$ leading to node $u$ has degree $k_i$, $[1-(1-x_i)^{k_i-1}]$ is the probability that at least one of the other $k_i-1$ outgoing connectivity links of node $u$ in network $i$ leads to the MCGC, and $\sum_{k_j} P_j(k_j)[1- (1-x_j)^{k_j}]$ is the probability that at least one of the $k_j$ connectivity links of node $w$ in network $j$ (node $w$ and node $u$ are connected by dependency link) leads to the MCGC.
 
Similarly extending En~\ref{En:double_mu}, we obtain the probability that a randomly selected node is in the MCGC as:
 
 \begin{equation}
 \mu^\infty=\prod_{n=1}^N \{ \sum_{k_n} P_n(k_n)[1- (1-x_n)^{k_n}]\}.
 \label{En:N_mu}
 \end{equation}
 
In the percolation problem, if $1-p$ fraction of nodes are randomly removed from layer $i$, we can simply multiply En~\ref{En:N_x} and \ref{En:N_mu} by $p$, which is the fraction of nodes remaining in the layer  after the attack:

 \begin{align}
 x_i&=p \cdot \sum_{k_i} \frac{P_i(k_i)k_i}{\langle k_i \rangle}[1-(1-x_i)^{k_i-1}] \prod_{j\neq i} \{\sum_{k_j} P_j(k_j)[1- (1-x_j)^{k_j}]\}
 \\
 \label{NX}
 &=F_i(p, x_1,x_2,\cdot \cdot \cdot ,x_N),
\\
 \mu^\infty&=p\cdot \prod_{n=1}^N \{ \sum_{k_n} P_n(k_n)[1- (1-x_n)^{k_n}]\}.
 \label{NMU}
 \end{align}
 
 \bigskip

{\it Example with Random Regular networks}
\bigskip
 
For a simple illustration, we let the networks to have the same degree distribution:
\begin{equation}
P_i(3)=P_j(3)=1,
\end{equation}
i.e. every network is a random regular network with degree 3.
 
  Since the equations are symmetric, and there is a one-to-one matching between every node on each network, we have the relation
 
\begin{equation}
x_1=x_2=\cdot \cdot \cdot =x_N=x.
\end{equation}

Now En~\ref{NX} simplifies into
\begin{equation}\label{ExampleRRXN}
x=p [1-(1-x)^2] [1- (1-x)^3]^{N-1}
\end{equation}
 Note that when $N=2$, we get exactly the En~\ref{ExampleRRX0} with $x=y$.
 
 By bringing $x$ to the right hand size, we could transform the En~\ref{ExampleRRXN} into
 \begin{equation}\label{ExampleRRXN0}
 F_x=p [1-(1-x)^2] [1- (1-x)^3]^{N-1}-x=0.
 \end{equation}
In this case, the critical value of $p_c$ can be understood as the smallest value of $p$ such that En~\ref{ExampleRRXN0} has a real solution of $x$ in the meaningful range of $[0,1]$. This means when $p<p_c$, En~\ref{ExampleRRXN0} only has a trivial solution of $x=0$, when $p>p_c$, there is more than one solution of $x>0$, and when $p=p_c$ one unique solution of $x$ exist .
 Thus at the critical point $x=x_c, p=p_c$, we would have the following relation fulfilled:
\begin{equation}
\frac{dF_x}{dx}|_{x=x_c} = 0,
\end{equation}
which leads to
\begin{equation}\label{ExampleRRXN1}
p\{2(1-x)[1-(1-x)^3]^{N-1}+3(N-1)(1-x)^2[1-(1-x)^3]^{N-2}\}-1=0.
\end{equation}
Solving the simultaneous equations \ref{ExampleRRXN0} and \ref{ExampleRRXN1} numerically, we are able to find out the value of $p^I_c$. Notice that for any integer value of $N>0$, we have a solution of $p_c$ and $x_c$  in the range of $[0,1]$, thus we always have a first order phase transition, but no second order one.

\section{Percolation on multi-layer interdependent networks with degree-degree correlations}

Usually in real-world networks, the connection through dependency links may not be random \cite{RoniEPL2010,SergeyPRE2011,BaxterPRL2012,YanqingPRE2013,LidiaPRETriplePoint2013,YanqingNP2014,GohPRECorr2014,GohNJPCorrMulti2012}. In a general form, we can assume a joint probability $P(k_A,k_B)$ for the dependency links to connect a node $u$ with degree $k_A$ in network $A$ and a node $u'$ with degree $k_B$ in network $B$. In this case, we still assume that each node is connected with one and only one node in the other network through a dependency link.

Instead of using the independent probabilities $P_A(k)$ and $P_B(k)$, the joint probability $P(k_A,k_B)$ is used. Thus En~\ref{STX}, \ref{STY} and \ref{STMU} become: 

\begin{equation}\label{CorrX}
x=p\cdot \sum_{k_A} \sum_{k_B}\frac{k_A}{\langle k_A \rangle}P(k_A,k_B) [1-(1-x)^{k_A-1}][1-(1-y)^{k_B}]
\end{equation}

\begin{equation}\label{CorrY}
y=p\cdot \sum_{k_B} \sum_{k_A}\frac{k_B}{\langle k_B \rangle} P(k_B,k_A)[1-(1-y)^{k_B-1}][1-(1-x)^{k_A}]
\end{equation}

\begin{equation}\label{CorrMU}
\mu^\infty=p\cdot \sum_{k_A} \sum_{k_B}P(k_A,k_B) [1-(1-x)^{k_A}][1-(1-y)^{k_B}]
\end{equation}

In fact, En~\ref{CorrX} \ref{CorrY} and \ref{CorrMU} are the more general representation of \ref{STX}, \ref{STY} and \ref{STMU}. 
In the case of Ref~\cite{SergeyPRE2011}, there is perfect correlation between the degrees of the two networks, i.e. $P(k_A,k_B)=P(k)$ if $k_A=k_B=k$; else $P(k_A,k_B)=0$. The above equations transform into:

\begin{equation}
x=p\cdot \sum_{k}\frac{k}{\langle k \rangle}P(k) [1-(1-x)^{k-1}][1-(1-y)^{k}]
\end{equation}

\begin{equation}
y=p\cdot \sum_{k}\frac{k}{\langle k \rangle} P(k)[1-(1-y)^{k-1}][1-(1-x)^{k}]
\end{equation}

\begin{equation}
\mu^\infty=p\cdot \sum_{k}P(k)[1-(1-x)^{k}][1-(1-y)^{k}]
\end{equation}

A special case is when we have random regular networks for both $A$ and $B$, and the results was discussed in the previous section since $P(k_A,k_B)=P(k_A)=P(k_B)=1$.

The more general case of correlated systems of a multiplex network with different types of links were studied in Ref~\cite{BaxterPRL2012}. With similar argument, in the case of correlated N-layer interdependent networks, we could write down the equations of $x_i$ and $\mu^\infty$ from En~\ref{NX} and \ref{NMU}:
\begin{align}
 x_i&=p \cdot \sum_{k_1,k_2,...} P(k_1,k_2,...) \frac{k_i}{\langle k_i \rangle}[1-(1-x_i)^{k_i-1}] \prod_{j\neq i} [1- (1-x_j)^{k_j}]
 \\
 \label{NXP}
 &=F_i(p, x_1,x_2,\cdot \cdot \cdot ,x_N),
\\
 \mu^\infty&=p\cdot  \sum_{k_1,k_2,...} P(k_1,k_2,...) \prod_{i=1}^N[1- (1-x_i)^{k_i}].
 \end{align}
Ref~\cite{BaxterPRL2012} provided a general mathematical tool to solve for the critical points by using the Jacobian of the equations:
\begin{equation}
det[\mathbf{J}-\mathbf{I}]=0,
\label{CriticalCondition}
\end{equation}
where $\mathbf{J}$ is the  Jacobian matrix with $J_{ij}=\partial F_i/\partial x_j$.
Solving En~\ref{NXP} and \ref{CriticalCondition} gives the critical point value of $p^I_c$, and $x_c$ for each layer of network in the system.

\section{Percolation on two layer partially interdependent networks}
In certain interdependent networks, not every node has a dependency link. It is more realistic to assume only a fraction of nodes from each network to have dependency links \cite{RoniPRL2010,LidiaPRETriplePoint2013}. And in such systems, both first and second order phase transitions may occur depending on the details of the networks' structural properties.

Let us assume two networks $A$ and $B$ with degree distribution $P_A(k)$ and $P_B(k)$, and only a fraction of $q$ nodes from each network is connected to nodes in the other network with dependency networks. For simplicity, we let each node to be connected with at most one other node through an dependency link.

In order for a randomly selected link in $A$ to lead to the MCGC, it must satisfy two conditions. First the node $u$ it directly attaches to must have at least one of its outgoing connectivity links leading to the MCGC, and the probability is the same as the case of full dependency links given by $\sum_k \frac{P_A(k)k}{\langle k \rangle } [1- (1-x)^{k-1}]$. Secondly, for the case of network $B$, there are two scenarios: there is a probability $1-q$ that node $u$ is not connected with any node in $B$, then $u$ is in the MCGC; there is a probability $q$ that $u$ is connected with a node $u'$ in $B$, then at least one of the connectivity links of $u'$ must also lead to the MCGC, and the probability for this is $q\sum_{k'}P_B(k')[1-(1-y)^{k'}]$. Therefore the original En~\ref{STX} becomes:

\begin{equation}
\label{PSTX}
x=p\cdot \sum_k \frac{P_A(k)k}{\langle k \rangle } [1- (1-x)^{k-1}]\cdot \{(1-q)+q\sum_{k'}P_B(k')[1-(1-y)^{k'}]\}.
\end{equation}
The parameter $p$ on the right hand side takes into account that after removal of $1-p$ nodes in network $A$ in the beginning of the attack, only a fraction of $p$ node remain.

It is worth noting that the calculation of $y$ is not symmetric with $x$, because there is no one-to-one matching between a node in $A$ and a node in $B$. For a node $u'$ in $B$, the difference is in the case when it has a dependency node $u$ in $A$ (probability $q$), at least one of the connectivity links of $u$ must lead to the MCGC (probability $\sum_{k'}P_A(k')[1-(1-x)^{k'}]$), and $u$ must not have been removed (probability $p$). Thus En~\ref{STY} becomes:
\begin{equation}\label{PSTY}
y=\sum_k \frac{P_B(k)k}{\langle k \rangle } [1- (1-y)^{k-1}]\cdot \{(1-q)+pq\sum_{k'}P_A(k')[1-(1-x)^{k'}]\}.
\end{equation}
There is no additional parameter $p$ in front of the right hand side since we are not removing nodes from network $B$ in the beginning.

Again, due to the lack of symmetry in this case, the sizes of the MCGC in $A$ and $B$ are expressed differently:
\begin{align}
\mu_A^\infty &= p\cdot \sum_k P_A(k) [1- (1-x)^k]\cdot \{(1-q)+q\sum_{k'}P_B(k')[1-(1-y)^{k'}]\}
\\
\mu_B^\infty&=\sum_k P_B(k) [1- (1-y)^k]\cdot \{(1-q)+pq\sum_{k'}P_A(k')[1-(1-x)^{k'}]\}
\end{align}

\subsubsection{Example with random regular networks}

For a simple illustration, we use random regular networks for both $A$ and $B$, and $P_A(3)=P_B(3)=1$.  Equations \ref {PSTX} and \ref {PSTY} simplifies into
 \begin{align}\label{Sx}
 x&=p[2x-x^2]\{1-q+q[1-(1-y)^3]\},\\
\label{Sy}
 y&=[2y-y^2]\{1-q+pq[1-(1-x)^3]\}.
 \end{align}
Note that, the cascading dynamics is not symmetric for network $A$ and $B$ for we only attack network $A$. For example,  when $q=0$, there is no dependency link between $A$ and $B$; If we remove all of the nodes in network $A$, none of the nodes in $B$ is affected and giant component still exist in $B$.  Without the loss of generality, we study the case where $1-p$ nodes are removed from $A$, and focus on the phase transition behavior in $A$.

For the second order phase transition of network A, $x=0$ at the critical point $p=p_c^{II}$. This means when $p\rightarrow p_c^{II}$, we have $x\rightarrow0$. Note that This does not imply $y\rightarrow0$ (we use $y_0$ denote this non-zero solution at the critical point). When $x\rightarrow0$, from En~\ref{Sy}  we have
 \begin{equation}
 y_0=2-\frac{1}{1-q}.
 \end{equation}
 Only the largest solution of $y_0$ in the range [0,1] is the realistic solution, thus $q$ must be in the range [0,0.5] for second order phase transition to occur. Usually for the more general cases, we could numerically get the solution of $y_0$ by iterative calculations starting from a value close to 1. Note however, this solution only depends on $q$. Submitting $y_0$ to Eq. \ref {Sx} and ignoring $x^2$ when $x\rightarrow0$, we have
\begin{equation}
 1=p_c^{II}2\{1-q+q[1-(1-y_0)^3]\}.
\end{equation}
Thus we can obtain the critical point value of $p$ for second order phase transition: 
\begin{equation}\label{pc22}
p_c^{II}=\frac{1}{2[1-q(1-(\frac{q}{1-q})^3])}.
\end{equation}

For the first order phase transition of network A, again we cannot assume $x\rightarrow 0$ as in the case of second order phase transition. Instead, by transforming Eqs. \ref {PSTX} and \ref {PSTY} we have 
\begin{equation}\label{PSTX1}
x=F_1(p, y)=2-\frac{1}{p\{1-q+q[1-(1-y)^3]\}},
 \end{equation}
and
 \begin{equation}\label{PSTY1}
y=F_2(p, x)=2-\frac{1}{1-q+pq[1-(1-x)^3]}.
 \end{equation} 
As argued earlier, the critical point value $p_c^I$ satisfies tangential requirement:
  \begin{equation}
 \frac{\partial F_1(p_c^I,y)}{\partial y} \cdot \frac{\partial F_2(p_c^I,x)}{\partial x}=1,
 \end{equation}
which can be written explicitly as
  \begin{equation}\label{TA11}
  \frac{3q(y-1)^2}{p_c^I\{q[(y-1)^3+1]-q+1\}^2}\cdot\frac{3p_c^Iq(x-1)^2}{\{p_c^Iq[(x-1)^3+1]-q+1\}^2}=1
  \end{equation}
Solving Eqs. \ref{Sx}, \ref{Sy} and \ref{TA11} for $x$, $y$ and $p_c^I$, we are able to get the threshold value $p_c^I$ numerically.

\begin{figure}
\includegraphics[scale=0.5]{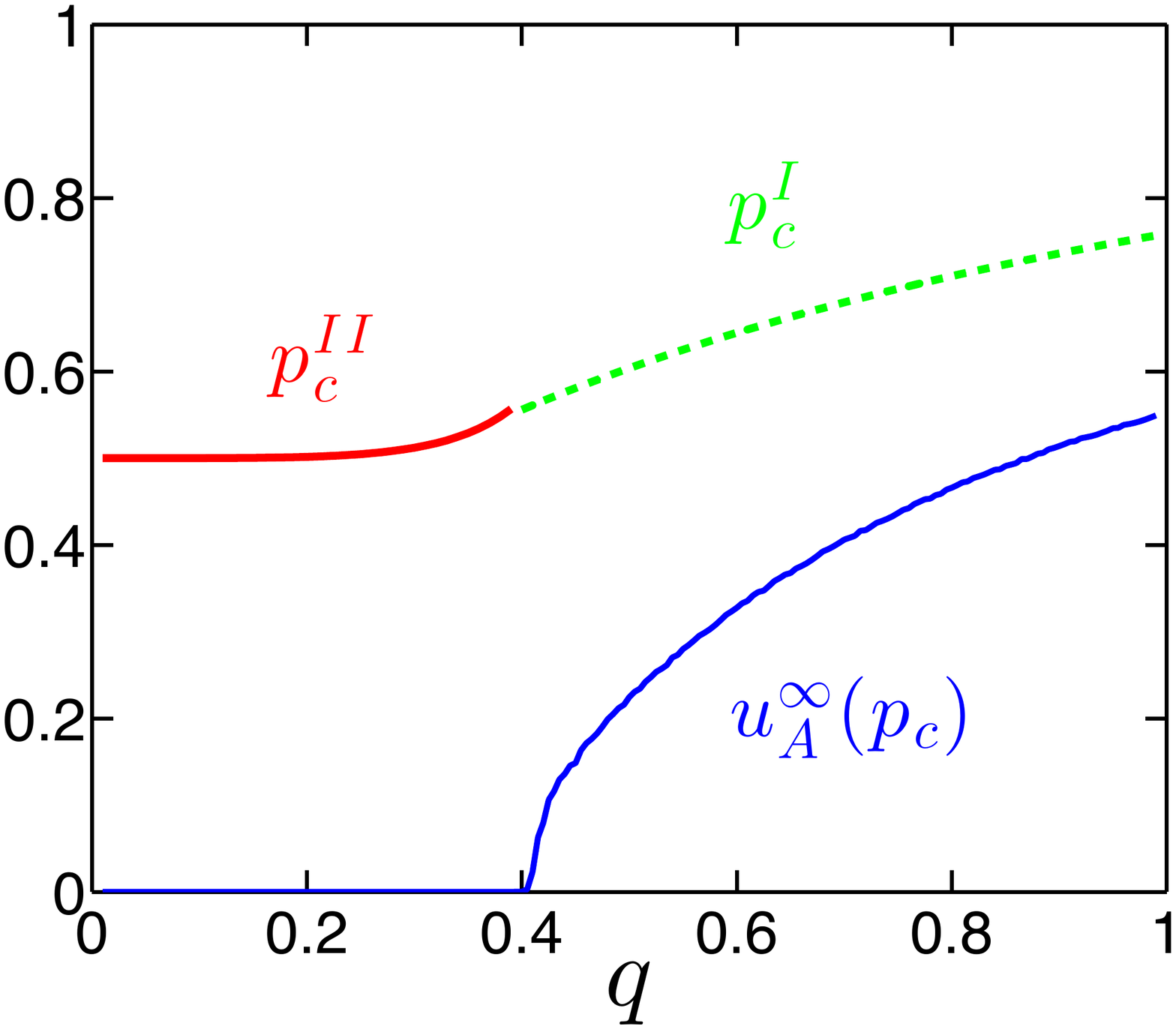}
\caption{\label{pc1pc2} $p_c^I$, $p_c^{II}$ and the giant component size $\mu_A^{\infty}(p_c)$ of network $A$ at the critical point. We have $p_c^I=p_c^{II}$ and $\mu_A^{\infty}(p_c)=0$ when $q=0.4$.  $\mu_A^{\infty}(p_c)>0$  and increases with $q$ when $q>0.4$.  It means that $q=0.4$ is the boundary of the second and first order phase transition. There is second order transition when $q<0.4$ and first order transition when $q>0.4$. $q=0.4$ is the boundary point for the system.}
\end{figure}

For any given value of $q$, which determines the fraction of dependency between networks $A$ and $B$, we would have either a first order phase transition with critical threshold $p_c^I$, or a second order phase transition with critical threshold $p_c^{II}$. Figure~\ref{pc1pc2} shows the plot of $p_c^I$ and $p_c^{II}$ v.s. the change in $q$. The two curves of $p_c^I$ and $p_c^{II}$ intersect at $q=0.4$. Hence the percolation behavior is separated into two regions: When 
$0\leq q\leq0.4$, it is second order transition; When
 $0.4<q\leq1$, it is first order  transition.

\subsubsection{Example with ER networks}

For a more general  example, we suppose the degree distributions of network $A$ and $B$ are both poisson with average degree $\langle k_A \rangle$ and $\langle k_B \rangle$. In this case we can use the generating function formulation \cite{NewmanPRE2001} to simplify the expressions. 

For networks $A$ and $B$, the corresponding generating functions are $G_0^A(x)=\sum P_A(k)x^{k}=e^{\langle k_A \rangle(x-1)}$ and $G_0^B(y)=\sum P_B(k)y^{k}=e^{\langle k_B\rangle(y-1)}$. The generating functions relating to the branching process are $G_1^A(x)=\sum \frac{P_A(k)k}{\langle k_A \rangle}x^{k'}=e^{\langle k_A \rangle(x-1)}$ and $G_1^B(y)=\sum \frac{P_B(k)k}{\langle k_B \rangle}y^{k}=e^{\langle k_B\rangle(y-1)}$. Note that the generating functions of degree distribution and the  the branching process are the same for poison degree distributions.

Thus, the equations \ref {PSTX} and \ref {PSTY} can be written as
 \begin{equation}\label{Sxx}
 x=p(1-e^{-\langle k_A \rangle x})[1-q+q(1-e^{-\langle k_B \rangle y})],
 \end{equation}
and
 \begin{equation}\label{Syy}
 y=(1-e^{-\langle k_B \rangle y})][1-q+pq(1-e^{-\langle k_A \rangle x})].
 \end{equation}
 Again, since to the nature of poisson degree distributions, the generating functions of degree distribution and the  the branching process are the same, we have
 $\mu_A^{\infty}=x$ and $\mu_B^{\infty}=y$. However, for other degree distributions this relation is generally invalid.
 
En~\ref{Sxx} and \ref{Syy} lead to the the explicit formula
 \begin{equation}\label{PSTX1}
x=F_1(p, y)=-\frac{1}{\langle k_A \rangle}\log(1+\frac{1-q}{pq}-\frac{y}{pq(1-e^{-\langle k_B \rangle y})}),
 \end{equation}
and
 \begin{equation}\label{PSTY1}
y=F_2(p, x)=-\frac{1}{\langle k_B \rangle}\log(\frac{1}{q}-\frac{x}{pq(1-e^{-\langle k_A \rangle x})}).
 \end{equation} 
 
 For the second order phase transition of network A, $x\rightarrow 0$ at the critical point. Using Eq. \ref{Syy} we have
\begin{equation}
 y_0=(1-q)(1-e^{-\langle k_B \rangle y_0}).
 \end{equation}
 
 Submitting $y_0$ to Eq. \ref{PSTY1}, we can obtain the explicitly formula of the second order phase transition critical point
  \begin{equation}\label{pc2}
  p_c^{II}=\frac{1}{\langle k_A \rangle[1-qe^{-\langle k_B \rangle y_0}]}.
 \end{equation}
 Here we use the first order term in the Taylor expansion of $e^{-\langle k_A \rangle x}$.
 
 For the first order phase transition of network A, using the tangential attachment of Eqs. \ref {PSTX1} and \ref {PSTY1} we have
 
 \begin{equation}\label{TA111}
 \frac{\partial F_1(p_c^I,y)}{\partial y} \cdot \frac{\partial F_2(p_c^I,x)}{\partial x}=1.
 \end{equation}
 
Again, Eqs. \ref{Sxx}, \ref{Syy} and \ref{TA111} allow us obtain $p_c^I$ numerically. Similar to the previous example, we can find out the value of $q$ such that 
\begin{equation} p_c^I=p_c^{II}. \end{equation} 
This would allow us to find the boundary between first and second order phase transitions, i.e. the triple point value.

Usually, the above three equations Ens.  \ref{PSTX1}, \ref{PSTX1} and \ref{TA111} have no explicit formula allow us to solve it directly. In order to detect $p_c^I$, we could brut-search for $p \in [0,1]$ according to following calculations. For a given $p$, we run Ens.  \ref{PSTX1} and \ref{PSTY1} iteratively and obtain the solutions (fixed point). Then put this fixed point $x, y$ into En. \ref{TA111}. If En. \ref{TA111} equal to 1, this $p$ should be $p_c^I$. For many more general case, such as the degree distribution is scale free. we have no explicit formula for En. \ref{TA111}, then you have to use numerical way to degree partial derivative at the fixed point $x, y$.

\begin{figure}
\includegraphics[scale=0.6]{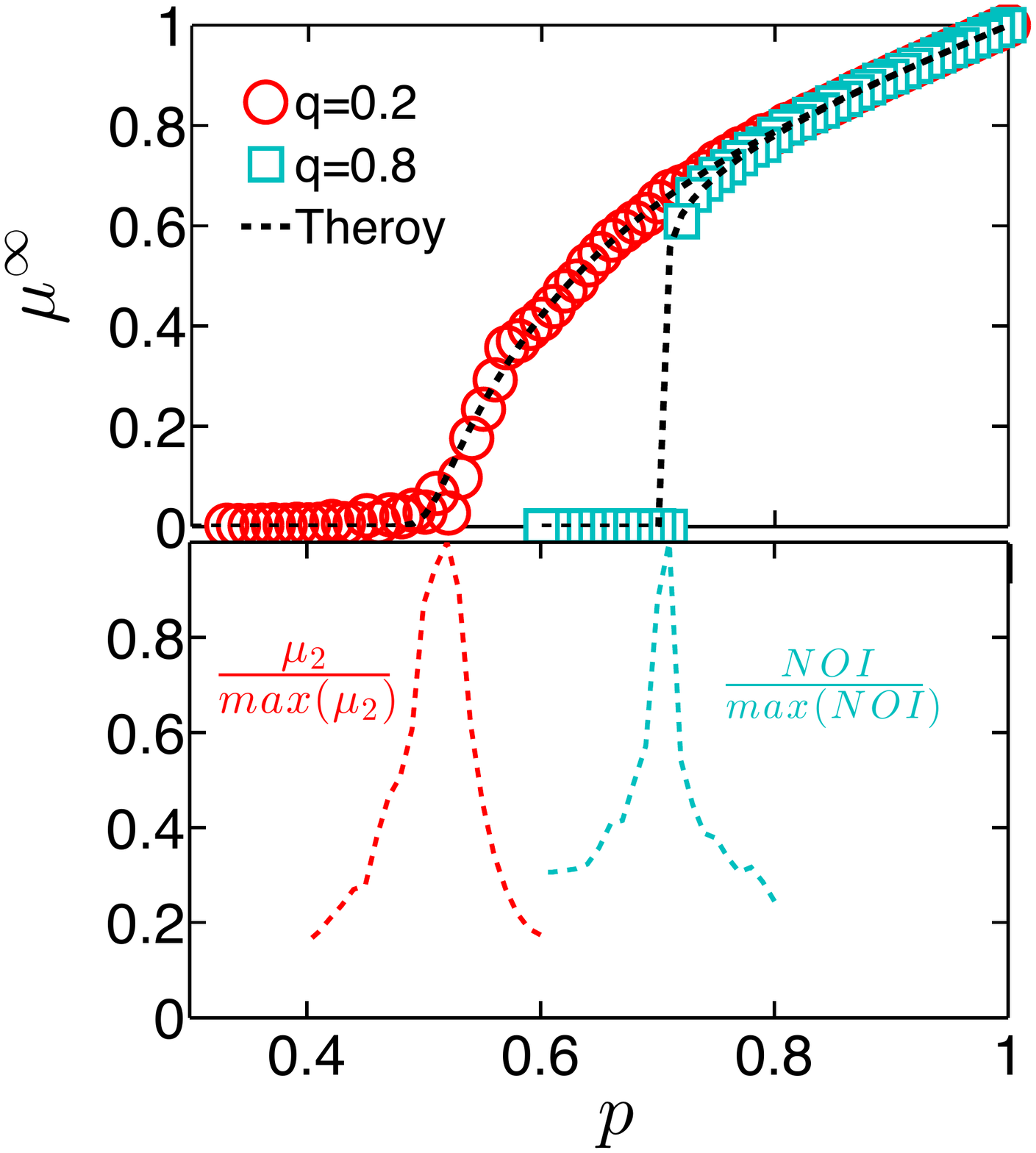}
\caption{\label{detectcriticalpoint} Identification of critical point through simulations. For 2nd order phase transitions (where $q=0.2$), the maximum size of the giant cluster appears at the critical point $p_c$. For 1st order phase transitions (where $q=0.8$), the number of cascading iterations is maximized at the critical point $p_c$.}
\label{simuCritical}
\end{figure}

Numerical simulation could help us to find the critical points without solving the equations \cite{RoniPNAS2011}. As demonstrated in figure~\ref{simuCritical}, for 2nd order phase transitions, the second largest cluster size $\mu_2$ is maximum at the critical point; repeated simulation for different value of $p$ can be carried out to find out the peak $\mu_2$ to identify $p_c$.
 For 1st order phase transitions, the number of iterations (NOI) is at maximum; Thus one can identify the $p_c$ as the point where maximum NOI is located. Here iteration refers to the cascading of failures from one network to the other.

\section{Single network with different types of links}

Dependency links could also exist in single layer networks~\cite{RoniPNAS2011,AmirPRE2011,YanqingPRX2014}, and the form of dependency could vary. This means, while the nodes in a single network are generally connected via connectivity links, some of the the nodes have mutual dependencies.

Suppose for a given network, certain pairs of nodes are mutually dependent on each other \cite{RoniPNAS2011}.  In this case, if node $u$ is dependent on node $w$, $w$ must and only depends on node $u$. 
Since the dependent nodes are in the same network, we could get the equation of $x$ simply by replacing $y$ with $x$ in En~\ref{STX} of the two interdependent networks.:
\begin{equation}
x=\{p \sum\frac{kP(k)}{\langle k \rangle}[1-(1-x)^{k-1}]\} \cdot \{ p\sum P(k)[1-(1-x)^k]\}.
\end{equation}
Note that the second $p$ on the right side is there because both the node itself and its dependent node have the probability $p$ to remain after the initial attack for they are in the same single network. Correspondingly, we can write down the size of the giant component as:
\begin{equation}
\mu^\infty=\{ p\sum P(k)[1-(1-x)^k]\}^2
\end{equation}

\begin{figure}[h]
 \includegraphics[width=0.5\linewidth]{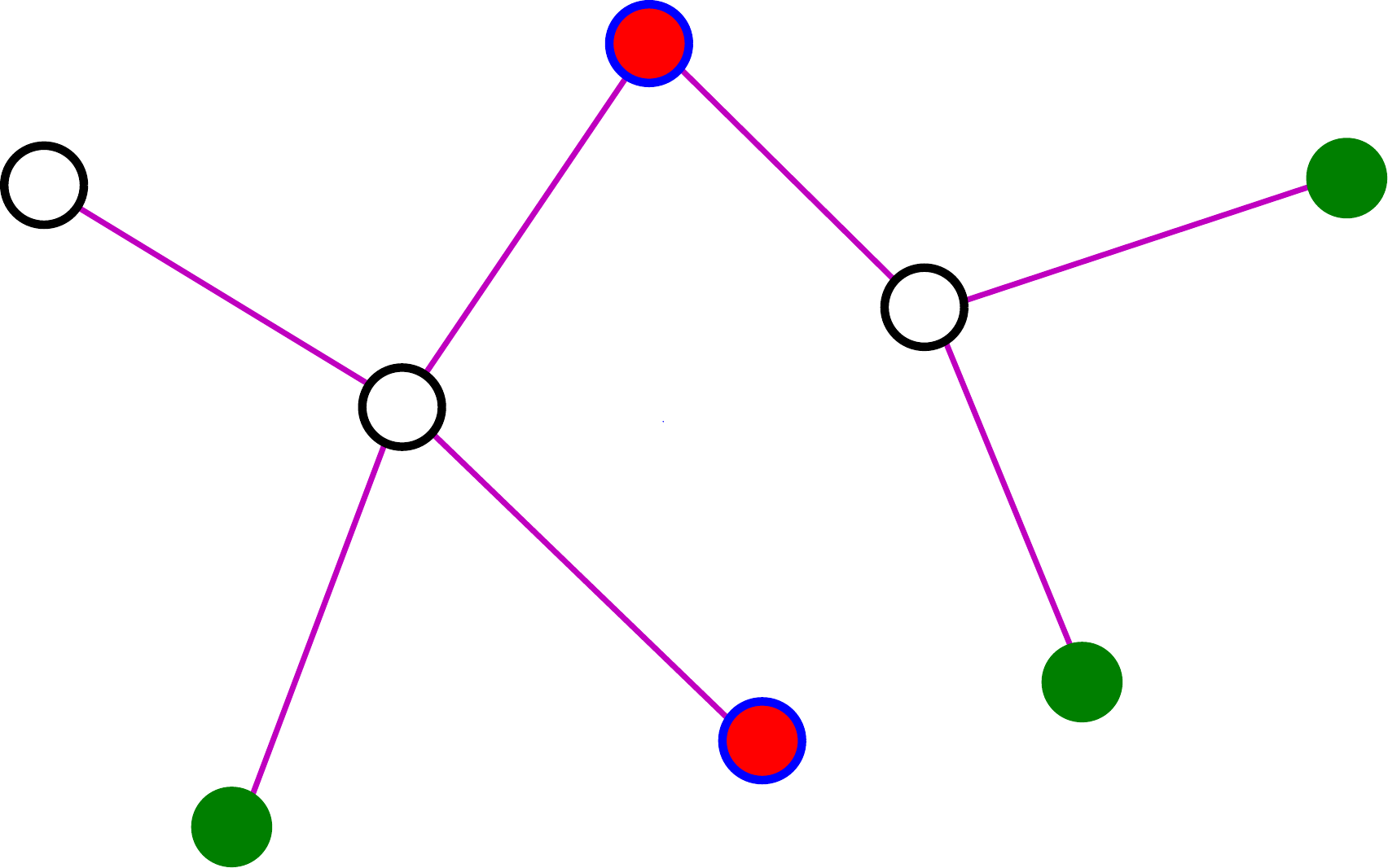}
 \caption{Single network with dependency groups. In this network, nodes in the same color are in the same dependency group. For example, if the green node on the bottom left is removed from the network, both of the two green nodes on the right would be removed due to their dependency relationship with the bottom left green node.}
\label{FIG:DependencyGroup}
\end{figure}

For a more general case, a network could have dependency groups \cite{AmirPRE2011} - certain group of nodes have dependency relations (as shown in Fig. \ref{FIG:DependencyGroup}) such that the removal of any one node would result in the removal of all the other nodes in the group. Given the probability distribution of the group size $g(s)$ where $s$ is the number of nodes in a group, we would obtain the following equation of $x$:
\begin{equation}
x=\{p \sum_k \frac{kP(k)}{\langle k \rangle}[1-(1-x)^{k-1}]\} \cdot  \sum_s \frac{s\cdot g(s)}{\langle s \rangle}\{ p\sum_k P(k)[1-(1-x)^k]\}^{s-1}.
\end{equation}
Note that in this case, every node in the dependency group of size $s$ needs to be in the giant component in order for the group to be in the giant cluster (else the whole dependency group would be removed). Here $\frac{s\cdot g(s)}{\langle s \rangle}$ is the probability that a node is a group of size $s$, and $\{ p\sum_k P(k)[1-(1-x)^k]\}^{s-1}$ is the probability that every other $s-1$ nodes is in the giant cluster. Using similar arguments, we could write down the equation for $\mu^\infty$:

\begin{equation}
\mu^\infty=\sum_s \frac{s\cdot g(s)}{\langle s \rangle} \{ p\sum_k P(k)[1-(1-x)^k]\}^s.
\end{equation}
 
From here, we could study the critical phase transition behaviors by solving the equations using the techniques in the previous sections.

\section{Percolation on two layer with many correlated links}

In certain networks including brain networks, it is observed that the dependency links are not always one-to-one matching, but could be one-to-many, and extensive correlations and dependencies between nodes exist. Ref~\cite{YanqingNP2014} discovered that brain networks are wired in such a way that stability is maximized.

To examine the critical phase transition behaviors of such networks, additional parameters need to be defined for specifying the structure. Assuming two networks $A$ and $B$, we denote $k_{in}^A$ ($k_{in}^B$) as the degree of connectivity links of a node in $A$($B$), and $k_{out}^A$ ($k_{out}^B$) as its dependency degree that lead to nodes in the other network. The joint probability that a node $u$ in $A$ has  $k_{in}^A$ connectivity links and $k_{out}^A$ dependency links is denoted by $p_A(k^A_{in},k^A_{out})$. The conditional probability $p_{AB}(k_{in}^B|k_{in}^A)$ is that given a node $u$ in $A$ with degree $k_{in}^A$, the probability that any of its dependent node $w$ in $B$ is of degree $k_{in}^B$. Similar definitions carry over to network B for $p_B(k^A_{in},k^A_{out})$ and $p_{AB}(k_{in}^A|k_{in}^B)$.
During an attack, a fraction of $1-p_1$ nodes in $A$ and $1-p_2$ nodes in $B$ are removed. This is in contrary to our previous examples in which only one network is being attacked at the beginning.

Here we let $x_A$($x_B$) be the probability that on following an arbitrary connectivity link in network $A$($B$), we reach a node leading to the giant component. For an dependency link between node $u$ with connectivity degree $k_{in}^A$ and node $w$ in $B$, $y_{k_{in}^A}$ is defined as the probability that  this link from $u$ leads to the giant component, and $y_{k_{in}^B}$ is defined similarly for a dependency link from $B$ to $A$. Thus we have the following self-consistent equations:

\begin{align}
x_A=p_1 \cdot \sum_{k^A_{in},k^A_{out}} \{\frac{p_A(k^A_{in},k^A_{out}) k^A_{in}}{\langle k^A_{in} \rangle}[1-(1-x_A)^{k^A_{in}-1}]\cdot [1-(1-y_{k_{in}^A})^{k^A_{out}}]\},
\\
x_B=p_2 \cdot \sum_{k^B_{in},k^B_{out}} \{\frac{p_B(k^B_{in},k^B_{out}) k^B_{in}}{\langle k^B_{in} \rangle}[1-(1-x_B)^{k^B_{in}-1}]\cdot [1-(1-y_{k_{in}^B})^{k^B_{out}}]\},
\end{align}
where $\frac{p_A(k^A_{in},k^A_{out}) k^A_{in}}{\langle k^A_{in} \rangle}$ is the probability that a randomly chosen link  in network $A$ leading to node $u$ has connectivity degree $k_{in}^A$ and dependency degree $k_{out}^A$, $[1-(1-x_A)^{k^A_{in}-1}]$ is the probability that at least one of the other $k_{in}^A-1$ outgoing connectivity links of node $u$ leads to the giant component, and $[1-(1-y_{k_{in}^A})^{k^A_{out}}]$ is the probability that at least one of the $k_{out}^A$ dependency links of $u$ leads to the giant component.

For the probability $y_{k_{in}^A}$ and $y_{k_{in}^B}$, we have
\begin{align}
y_{k_{in}^A}=p_2\{\sum_{k_{in}^B} p_{AB}(k_{in}^B|k_{in}^A)[1-(1-x_B)^{k_{in}^B}]\},
\\
y_{k_{in}^B}=p_1\{\sum_{k_{in}^A} p_{AB}(k_{in}^A|k_{in}^B)[1-(1-x_A)^{k_{in}^A}]\},
\end{align}
where $[1-(1-x_A)^{k_{in}^A}]$ is the probability that at least one of the $k_{out}^A$ dependency neighbors (in network $B$) of node $u$ (which has degree $k_{in}^A$) is in the giant component. Finally we can write down the probabilities that a randomly chosen node is in the giant component:
\begin{align}
\mu_A^{\infty}=p_1\{\sum_{k^A_{in},k^A_{out}} P_A(k^A_{in},k^A_{out})[1-(1-x_A)^{k^A_{in}}]\cdot[1-(1-y_{k_{in}^A})^{k^A_{out}}]\},
\\
\mu_B^{\infty}=p_2\{\sum_{k^B_{in},k^B_{out}} P_B(k^B_{in},k^B_{out})[1-(1-x_B)^{k^B_{in}}]\cdot[1-(1-y_{k_{in}^B})^{k^B_{out}}]\},
\end{align}
which are straight forward.

In general, since the above system is extremely complicated to have analytical solutions, numerical methods are usually preferred.  Here we illustrate a simple yet efficient numerical method called binary search to detect the critical point for network A. The same method can be easily applied to network B. First we setup the initial starting points with $p_{c-}=0$ and $p_{c^+}=1$, and let $p_c=\frac{p_{c^-}+p_{c^+}}{2}$. If $\mu_A^{\infty}(p_c)=0$, we change the value of $p_{c^-}$ by letting $p_{c^-}=p_c$; otherwise, we change $p_{c+}$ by letting $p_{c+}=p_c$. A $p_c$ value with high precision could usually be reached with 20 such iterations as this algorithm converges exponentially.

\section{Conclusion}

In this work, we have provided a specific mathematical framework to review the critical phase transition behavior of interdependent networks, otherwise known as Network of Networks (NON). Starting from single random networks, we have shown that by defining two key mathematical quantities - the probabilities of finding a link/node in the final giant component, - one is able to directly write down the sets of self-consistent equations of these quantities without going through the iterative process of cascading failures in stages. This methodology greatly simplifies the mathematical analysis in complicated  network structures, especially in very complex systems involving correlations and multiple dependency links per node.

There has been many other works we have not included here. For example, Ref~\cite{YanqingPRX2014} has analyzed multiplex directed networks in the context of social networks.  \cite{GomezSP2012, wang2012probabilistic,Wang2013SR} have studies evolutionary games. Interdependent networks with spatial constraint \cite{BashanNP2013,li2012cascading} have been shown to exhibit unique phase transition behaviors, though we have not discussed them due to their analytical difficulties. In this work, our focus  is  to provide an mathematical overview using this specific technique of simplified self-consistent probabilities. The recursive mapping method \cite{SergeyNature2010} yields the same results, but we demonstrated that this particular method could greatly simplify the mathematical derivations for a wide range of complicated systems.

Although this method proves to be applicable to a wide range of networks systems in studying their percolation behaviors, caution must be taken when implementing it. It is crucial that the self-consistent equations need to be carefully constructed, such that every component of the equations strictly follow the branching process underlying the percolation behaviors.

\section{Acknowledgement}
This work is partially supported by the NSFC  Grant No. 61203156 and the Fundamental Research Funds for the Central Universities Gran No. 2682014RC17.

\bigskip

\bibliographystyle{unsrt}

\end{document}